\PassOptionsToPackage{warn}{textcomp}

\documentclass[sigconf]{acmart}

\renewcommand\footnotetextcopyrightpermission[1]{} % removes footnote with conference information in first column

\usepackage{booktabs} % For formal tables

\usepackage{times}
\usepackage{latexsym}
\usepackage{tabularx}
\usepackage{graphicx}
\usepackage{mathtools}
\usepackage{url}
\usepackage{subfig}
\usepackage{supertabular,booktabs}
\usepackage{xcolor}
\usepackage{color, colortbl,soul}
\usepackage[utf8]{inputenc}
\usepackage[greek,english]{babel}
\usepackage{textgreek}

\definecolor{LGray}{gray}{0.9}
\definecolor{Gray}{gray}{0.8}
\definecolor{DGray}{gray}{0.7}
\definecolor{LBlue1}{rgb}{0.63, 0.79, 0.95}
\definecolor{LBlue}{rgb}{0.74, 0.83, 0.9}
\definecolor{LRed}{rgb}{0.99, 0.76, 0.8}

\copyrightyear{2018} 
\acmYear{2018} 
\setcopyright{acmlicensed}
\acmConference[CIKM '18]{2018 ACM Conference on Information and Knowledge Management}{October 22--26, 2018}{Torino, Italy}
\acmBooktitle{2018 ACM Conference on Information and Knowledge Management (CIKM'18), October 22--26, 2018, Torino, Italy}
\acmPrice{15.00}
\acmDOI{10.1145/3269206.3271783}
\acmISBN{978-1-4503-6014-2/18/10}

\begin{document}
\title{Nowcasting the Stance of Social Media Users in a Sudden Vote: The Case of the Greek Referendum}

%\subtitlenote{The full version of the author's guide is available as \texttt{acmart.pdf} document}

\author{Adam Tsakalidis}
%\authornote{Dr.~Trovato insisted his name be first.}
\affiliation{%
  \institution{University of Warwick}
  %\city{Coventry}
  %\country{UK}
}
\affiliation{\institution{The Alan Turing Institute}}
\email{atsakalidis@turing.ac.uk}

\author{Nikolaos Aletras}
%\authornote{The secretary disavows any knowledge of this author's actions.}
\affiliation{%
  \institution{University of Sheffield}
  %\streetaddress{P.O. Box 1212}
  %\city{Sheffield}
  %\country{UK}
}
\email{n.aletras@sheffield.ac.uk}

\author{Alexandra I. Cristea}
%\authornote{This author is the one who did all the really hard work.}
\affiliation{%
  \institution{Durham University}
  \institution{University of Warwick}
  %\streetaddress{1 Th{\o}rv{\"a}ld Circle}
  %\city{Durham}
  %\country{UK}
}
\email{alexandra.i.cristea@durham.ac.uk}

\author{Maria Liakata}
\affiliation{%
  \institution{University of Warwick}
  %\city{Coventry}
  %\country{UK}
}
\affiliation{\institution{The Alan Turing Institute}}
\email{m.liakata@warwick.ac.uk}

\renewcommand{\shortauthors}{A. Tsakalidis et al.}

\begin{abstract}
Modelling user voting intention in social media is an important research area, with applications in analysing electorate behaviour, online political campaigning and advertising. Previous approaches mainly focus on predicting national general elections, which are regularly scheduled and where data of past results and opinion polls are available. However, there is no evidence of how such models would perform during a sudden vote under time-constrained circumstances. That poses a more challenging task compared to traditional elections, due to its spontaneous nature. In this paper, we focus on the 2015 Greek bailout referendum, aiming to nowcast on a daily basis the voting intention of 2,197 Twitter users. We propose a semi-supervised multiple convolution kernel learning approach, leveraging temporally sensitive text and network information. Our evaluation under a real-time simulation framework demonstrates the effectiveness and robustness of our approach against competitive baselines, achieving a significant 20\% increase in F-score compared to solely text-based models.
\end{abstract}




\maketitle

\section{Introduction}

Predicting user voting stance and final results in elections using social media content is an important area of research in social media analysis \citep{metaxas2011not,gayo2012wanted} with applications in online political campaigning and advertising~\cite{Howard2005,Cogburn2011}. It also provides political scientists with tools for qualitative analysis of electoral behaviour on a large scale~\cite{Aldrich2016}. Previous approaches mainly focus on predicting national general elections, which are regularly scheduled and where data of past results and opinion polls are available \cite{lampos2013user,tsakalidis2015predicting}. However, there is no evidence of how such models would work during a \textit{sudden} and \textit{major} political event under time-constrained circumstances. That forms a more challenging task compared to general elections, due to its spontaneous nature \cite{leduc2002opinion}. Building robust methods for voting intention of social media users under such circumstances is important for political campaign strategists and decision makers.

Our work focuses on nowcasting the voting intention of Twitter users in the 2015 Greek bailout referendum that was announced in June, 27$^{th}$ 2015 and was held eight days later. We define a time-sensitive binary classification task where the aim is to classify a user's voting intention (\texttt{YES}/\texttt{NO}) at different time points during the entire pre-electoral period. 

For this purpose, we collect a large stream of tweets in Greek and manually annotate a set of users for testing. We also collect a set of users for training via distant supervision. We predict the voting intention of the test users during the eight-day period until the day of the referendum with a multiple convolution kernel learning model. The latter allows us to leverage both temporally sensitive textual and network information. Collecting all the available tweets written in Greek\footnote{As per Twitter Streaming API limitations: \url{https://developer.twitter.com/en/docs/basics/rate-limiting}}, enables us to study user language use and network dynamics in a complete way. We demonstrate the effectiveness and robustness of our approach, achieving a significant 20\% increase in F-score against competitive text-based baselines. We also show the importance of combining text and network information for inferring users' voting intention. 

Our paper makes the following contributions:

\begin{itemize}
    \item We present the first systematic study on nowcasting the voting intention of Twitter users during a sudden and major political event.
    
    \item We demonstrate that network and language information are complementary, by combining them with multiple convolutional kernels.
    
    \item We highlight the importance of the temporal modelling of text for capturing the voting intention of Twitter users.
    
    \item We provide qualitative insights on the political discourse and user behaviour during this major political crisis.    
\end{itemize}

\section{Related Work}

Most previous work on predicting electoral results focuses on forecasting the final outcome. Early approaches based on word counts \cite{tumasjan2010predicting} fail to generalize well \cite{metaxas2011not,gayo2012wanted,jungherr2012pirate}. \citet{lampos2013user} presented a bilinear model based on text and user information, using opinion polls as the target variable. \citet{tsakalidis2015predicting} similarly predicted the election results in different countries using Twitter and polls while others used sentiment analysis methods and past results \cite{Oconnor2010,Shi2012,Ceron2014,Burnap2016}. More recently, \citet{Swamy2017} presented a method to forecast the results of the latest US presidential election from user predictions on Twitter. The key difference between our task and this strand of previous work lies in its spontaneous and time-sensitive nature. Incorporating opinion polls or past results is not feasible, due to the time-constrained referendum period and the lack of previous referendum cases, respectively. Previous work on predicting the outcomes of referendums \cite{Celli2016,Grcar2017,Lopez2017} is also different to our task, since they do not attempt to predict a single user's voting intention but rather make use of aggregated data coming from multiple users to predict the voting share of only a few test instances.

On the user-level, most past work has focused on identifying the political leaning (left/right) of a user. 
Early work by \citet{rao2010classifying} explored the linguistic aspect of the task; follow-up work has also incorporated features based on the user's network \cite{pennacchiotti2011machine,al2012homophily,conover2011predicting,volkova2014inferring}, leading to improvements in performance. However, most of this work predicts the (static) political ideology of clearly separated groups of users who are either declaring their political affiliation in their profiles, or following specific accounts related to a political party. This has been demonstrated to be problematic when applying such models on users that do not express political opinion \cite{cohen2013classifying}. \citet{preoctiuc2017beyond} proposed instead a non-binary, seven-point scale for measuring the self-reported political orientation of Twitter users, showcasing that the task is more difficult for users who are not necessarily declaring their political ideology. Our work goes beyond political ideology prediction, by simulating a real-world setting on a dynamically evolving situation for which there is no prior knowledge.

A smaller body of research has focused on tasks that go beyond the classic left/right political leaning prediction. \citet{fang2015topic} predicted the stance of Twitter users in the 2014 Scottish Independence referendum by analysing topics in related online discussions.
%Other previous work has classified Twitter users based on their political party preference in the UK \cite{boutet2012s} and the U.S. \cite{pennacchiotti2011democrats}. 
In a related task, \citet{zubiaga2017stance} classified user stance in three independence movements while \citet{Stewart2018} analysed user linguistic identity in the Catalan referendum. Albeit relevant, none of these works have actually studied the problem under a real-time evaluation setting or during a sudden event where the time between announcement and voting day is extremely limited (e.g., less than two weeks). Previous work on social media analysis during the Greek bailout referendum \cite{michailidou2017twitter,antonakaki2017social} has not studied the task of inferring user voting intention, whereas most of the past work in opinion mining in social media in the Greek language has focused primarily on tasks related to sentiment analysis \cite{kalamatianos2015sentiment,palogiannidi2015affective,tsakalidis2018building}. To the best of our knowledge, this is the first work to (a) infer user voting intention under \textit{sudden} circumstances and a \textit{major} political crisis; and (b) model user information over time under such settings.

\section{The Greek Bailout Referendum}

The period of the Greek economic crisis before the bailout referendum (2009-2015) was characterized by extreme political turbulence, when Greece faced six straight years of economic recession and five consecutive years under two bailout programs~\cite{Tsebelis2016}. Greek governments agreed to implement austerity measures, in order to secure loans and avoid bankruptcy -- a fact that caused massive unrest and demonstrations. During the same period, political parties regardless of their side on the left-right political spectrum were divided into \emph{pro-austerity} and \emph{anti-austerity}, while the traditional two-party system conceived a big blow \cite{Bosco2012,Teperoglou2012,Rudig2013}. 

The Greek bailout referendum was announced on June, 27$^{th}$ 2015 and was held eight days later. The Greek citizens were asked to respond as to whether they agree or not (\texttt{YES}/\texttt{NO}) with the new bailout deal proposed by the Troika\footnote{A decision group formed by the European Commission, the European Central Bank and the International Monetary Fund to deal with the Greek economic crisis.} to the Greek Government in order to extend its credit line. The final result was 61.3\%-38.7\% in favor of the \texttt{NO} vote. For more details on the Greek crisis, refer to \citet{Tsebelis2016}.

\section{Task Description} 
\label{sec:taskdesc}

Our aim is to classify a Twitter user either as a \texttt{YES} or a \texttt{NO} voter in the Greek Bailout referendum over the eight-day period starting right before its announcement (26/6, day 0) and ending on the last day before it took place (4/7, day 8). 

We assume a training set of users $D_t=\{(x_{t}^{(1)},y^{(1)}),...,(x_{t}^{(n)},y^{(n)})\}$, where $x_{t}^{(i)}$ is a representation of user $i$ up to time step $t \in [0,...,8]$ and $y^{(i)} \in \{\texttt{YES},\texttt{NO}\}$. Given $D_t$, we want to learn a function $f_t$ that maps a user $j$ to her or his stance $\hat{y}^{(j)}=f_t(x_{t}^{(j)})$ at time $t$. Then, we update our model with new information shared by the users in our training set up to $t$+1, to predict the test users voting intention at $t$+1. %For example, when $t=2$, we predict the stance of users using the information in $[0,1,2]$ where 0 represents the information available up to the day before the announcement. 
Therefore, we mimic a real-time setup, where we nowcast user voting intention, starting from the moment before the announcement of the referendum, until the day of the referendum. Sections~\ref{sec:data} and \ref{sec:methods} present how we develop the training dataset $D_t$ and the function $f_t$ respectively.

%define a binary classification task in each time step  .

%In particular, we want to train a model on a labelled training set consisting of users and their tweets up to time $t$, in order to predict some test users' voting intention at time $t$. Then, we want to update our model with new data shared by the users in our training set up to $t$+1, in order to predict the test users' voting intention at $t$+1, based on the new information. This way, we mimic a real-world setting, in which we are monitoring public opinion, starting from the moment before the announcement of the referendum, until the day of the referendum. For that purpose we define 4 periods: (1) Before the announcement; (2) After the announcement and before the referendum; (3) During the day of the referendum; and (4) After the referendum.

\section{Data}
\label{sec:data}

Using the Twitter Streaming API during the period 18/6--16/7, we collected 14.62M tweets in Greek (from 304K users) containing at least one of 283 common Greek stopwords, starting eight days before the announcement of the referendum and stopping 11 days after the referendum date (see Figure~\ref{fig:dataset}). This provides us with a rare opportunity to study the interaction patterns among the users in a rather complete and unbiased setting, as opposed to the vast majority of past works, which track event-related keywords only. For example, \citet{antonakaki2017social} collected 0.3M tweets using popular referendum-related hashtags during  25/06--05/07 -- we have collected 6.4M tweets during the same period. In the rest of this section, we provide details on how we processed the data in order to generate our training set in a semi-supervised way (\ref{sec:51}) and how we annotated the users that were used as our test set in our experiments (\ref{sec:52}).

% for the 4 periods 

\begin{figure}[!t]
\centering
\includegraphics[width=.92\columnwidth]{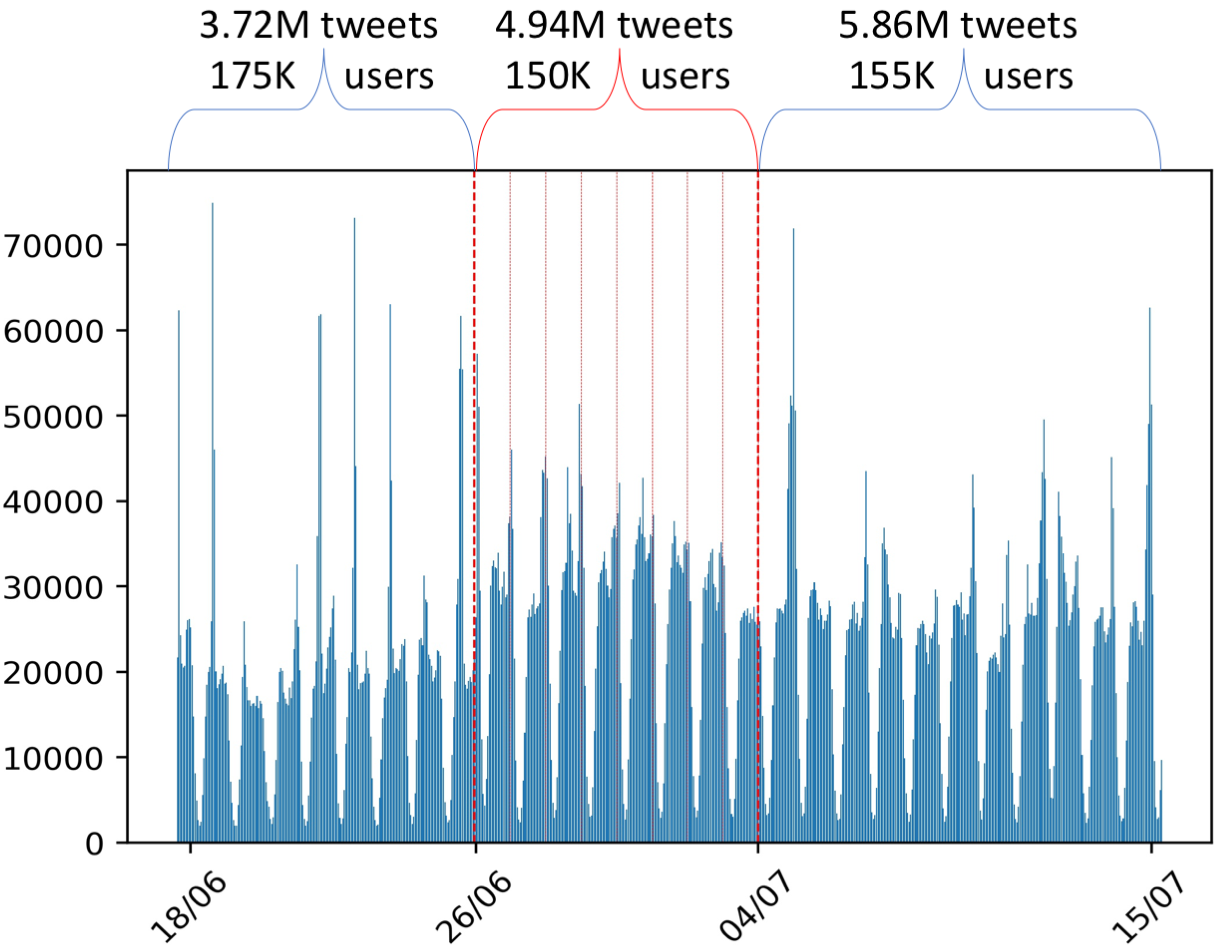}
\caption{Number of tweets in Greek per hour. The period highlighted in red indicates the nine evaluation time points (see Section~\ref{sec:taskdesc}).}%, starting before the announcement of the referendum and ending in the day before the date of the referendum.}
\label{fig:dataset}
\end{figure}

\iffalse
\begin{figure}[!t]
\centering
\includegraphics[width=\columnwidth]{numTweets2.jpg}
\caption{Number of tweets per hour.}
\label{fig:dataset}
\end{figure}

\begin{table}[!t]
\centering
\resizebox{\columnwidth}{!}{
\begin{tabular}{|l|l|l|r|r|}\hline
\#         & \textbf{{[}Start--End{]}}& \textbf{Description}& \textbf{\#tweets} & \textbf{\#users} \\ \hline
1& {[}18/06--26/06{]}& \begin{tabular}[c]{@{}l@{}}Before A \end{tabular}&3.72M&175K\\ \hline
2&{[}27/6--05/07{]}&\begin{tabular}[c]{@{}l@{}}Before R\end{tabular} &4.94M&150K\\ \hline
3&\begin{tabular}[c]{@{}l@{}}05/07 (07:00--19:00)\end{tabular}&\begin{tabular}[c]{@{}l@{}}During R\end{tabular}&0.27M&33K\\ \hline
4&{[}05/07--16/07{]}&\begin{tabular}[c]{@{}l@{}}{}After R\end{tabular}&5.69M&153K\\ \hline \hline
 & {[}18/06--16/07{]}  & Overall & 14.62M &  304K \\ \hline 
\end{tabular}}
\caption{Summary of our dataset divided into four time periods with respect to the  announcement of the referendum (A) and the referendum day (R).}\label{tab:dataset}
\end{table}
\fi

\subsection{Training Set} 
\label{sec:51}

Manually creating a training set would have required annotating users based on their voting preference on an issue that they had not been aware of prior to the referendum announcement. However, the same does not hold for certain accounts (e.g., major political parties) whose stance on austerity had been known a-priori given their manifestos and previous similar votes in  parliament~\cite{Rudig2013}. Such accounts can be used as seeds to form a semi-supervised task, under the hypothesis that users who are re-tweeting a political party more often than others, are likely to follow its stance in the referendum, once this is announced. Hence, we compile a set of 267 seed accounts (148 \texttt{YES}, 119 \texttt{NO}) focusing on the pre-announcement period including: (1) \emph{political parties}; (2) \emph{members of parliament (MPs)}; and (3) \emph{political party members}.
%\iffalse
\begin{itemize}
\item \textbf{Political Parties}:  We add as seeds the Twitter accounts of nine major and minor parties\footnote{We excluded KKE (Greek Communist Party) since an active official Twitter account did not exist at the time.} with a known stance on austerity before the referendum (5 \texttt{YES}, 4 \texttt{NO}, see Table~\ref{tab:parties}). We assume that the pro-austerity parties will back the bailout proposal (\texttt{YES}), while the anti-austerity parties will reject it (\texttt{NO}). The pro-/anti- austerity stance of the parties was known before the referendum, since the pro-austerity parties had already backed previous bailout programs in parliament or had a clear favorable stance towards them, whereas the opposite holds for the anti-austerity parties~\cite{Rudig2013}. %Therefore, we consider as seeds the Twitter accounts of all parties\footnote{We excluded KKE since they did not have an official Twitter account at the time.} with a known stance on austerity before the referendum (see Table ~\ref{tab:parties}).%; we added three minor political parties (ANTARSYA for \texttt{NO}, KIDISO and Dimiourgia Xana for \texttt{YES}), since those were active in online networks and had a clear stance on austerity\footnote{We excluded KKE since they did not have an official Twitter account at the time.}.

\item \textbf{MPs} The accounts of the (300) MPs of these parties were manually extracted and added as seeds. 153 such accounts were identified (82 \texttt{YES}, 71 \texttt{NO}) labelled according to the austerity stance of their affiliated party.

\item \textbf{Political Party Members} We finally compiled a set of politically related keywords to look up in Twitter user account names and descriptions (names/abbreviations of the nine parties and keywords such as ``candidate''). We identified 257 accounts (133 \texttt{YES}, 124 \texttt{NO}), which were manually inspected by human experts to filter out irrelevant ones (e.g., the word ``River'' might not refer to the political party) and kept only those that had at least one tweet during the period preceding the announcement of the referendum (44 \texttt{NO}, 61 \texttt{YES}).
\end{itemize}
%\fi

% \noindent\underline{\textit{-- Political parties}}:  We hypothesize that the pro-austerity/pro-EU parties will back the bailout proposal (YES vote), while the anti-austerity parties will reject it (NO vote). Pro-austerity parties had already backed previous bailout programs or had a clear stance on them, whereas the opposite holds for the anti-austerity parties. The Twitter accounts of the major political parties were divided into two classes, based on their stance on austerity measures (see Table ~\ref{tab:parties}); we added three minor political parties (ANTARSYA for \texttt{NO}, KIDISO/Dimiourgia Xana for \texttt{YES}), since those were active in online networks and had a clear stance on austerity\footnote{We excluded KKE since they did not have an official Twitter account at the time.}.

% \noindent\underline{\textit{--MPs}}: The Twitter accounts of the MPs of these parties were added as seeds in the corresponding class. 153 (out of 300 possible) such accounts were identified (71 for \texttt{NO}, 82 for \texttt{YES}). MPs accounts are labelled as YES or NO using the austerity stance of their affiliated party.
    
% \noindent\underline{\textit{--Political Accounts}}: We compiled a set of politically-related keywords to search for in the Twitter account names and descriptions (names/abbreviations of the nine parties and keywords such as ``candidate''). We identified 133/124 \texttt{NO}/\texttt{YES} accounts, which were inspected for sanity check (e.g., the word ``River'' is ambiguous).

\begin{table}[!t]
\renewcommand{\arraystretch}{1.2}
\centering
\caption{Political position, austerity, referendum stance and national election result (January 2015) of the political parties that are used as seeds in our modelling.}\label{tab:parties}
\resizebox{\columnwidth}{!}{
\begin{tabular}{l|l|l|l|c}\hline
\rowcolor{DGray} \textbf{Party}  & \textbf{Position} & \textbf{Auster.} & \textbf{Referend.} & \textbf{Jan 15 (\%)}\\ \hline
 SYRIZA (\textgreek{ΣΥΡΙΖΑ})            & Left               & anti   & NO  &  36.34    \\ \hline
\rowcolor{LGray} New Democracy (\textgreek{Νέα Δημοκρατία})     & Centre-Right       & pro    & YES &  27.81     \\ \hline
 Golden Dawn (\textgreek{Χρυσή Αυγή})        & Far-right          & anti   & NO  &  6.28   \\ \hline
\rowcolor{LGray} The River (\textgreek{Το Ποτάμι})          & Centre             & pro    & YES &  6.05      \\ \hline
 Independent Greeks (\textgreek{Ανεξάρτητοι Έλληνες}) & Right              & anti   & NO  &  4.75   \\ \hline
\rowcolor{LGray} PASOK (\textgreek{ΠΑΣΟΚ})  & Centre-left        & pro    & YES &  4.68       \\ \hline
 KIDISO  (\textgreek{ΚΙΔΗΣΟ})           & Centre-left        & pro    & YES &  2.47  \\ \hline
\rowcolor{LGray} ANTARSYA   (\textgreek{ΑΝΤΑΡΣΥΑ})        & Far-left           & anti   & NO  &  0.64   \\ \hline
 Creation Again (\textgreek{Δημιουργία Ξανά})   & Centre-Right       & pro    & YES &  - \\ \hline
\end{tabular}}
\end{table}

%We focused strictly on the pre-announcement period and found 119/148 \texttt{NO}/\texttt{YES} seeds, using the union of the aforementioned sets. 

To expand the set of seed accounts, we calculate for every user $u$ in our dataset during the pre-announcement period his/her score as:
\begin{equation*}
score(u) = PMI(u, \texttt{YES}) - PMI (u, \texttt{NO}) ,    
\end{equation*}
where $PMI(u, lbl)$ is the pointwise mutual information between a certain user and the respective seeding class (\texttt{YES}/\texttt{NO}). A high (low) score implies that the user is endorsing often \texttt{YES}-related (\texttt{NO}-related) accounts, thus he/she is more likely to follow their stance after the referendum is announced. This approach has been successfully applied to other related natural language processing tasks, such as building sentiment analysis lexical resources using a pre-defined list of seed words \cite{mohammad2013nrc}. Assigning class labels to the users based on their scores, we set up a threshold $tr = n(\max(|scores|))$, with $n \in [0,1]$. We assign the label \texttt{YES} to a user $u$ if $score(u)>tr$ or \texttt{NO} if $score(u)<-tr$.  Setting $n=0$, would imply that we are assigning the label \texttt{YES} if the user has re-tweeted more YES-supporting accounts (and inversely), which might result into a low quality training set, whereas higher values for $n$ would imply a smaller (but of higher quality) training set. During development, we empirically set $n=0.5$ to keep users who are fairly closer to one class than the other. From the final set of 5,430 users that have re-tweeted any seed account, 2,121 were kept (along with the seed accounts) as our training set (965 \texttt{YES}, 1,156 \texttt{NO}).%, with an overall number of 1,431,664 tweets in our full dataset.

\begin{table}[!t]
\centering
\caption{Number of users (u) and tweets (t) used in our experiments per evaluation day.}
\label{tab:testusers}
\resizebox{\columnwidth}{!}{
\begin{tabular}{l|ccccccccc}\hline
\rowcolor{DGray} day    & {\bf0}    & {\bf1}    & {\bf2}    & {\bf3}   & {\bf4}    & {\bf5}    & {\bf6}    & {\bf7}    & {\bf8}    \\\hline

date &26/06&27/06&28/06&29/06&30/06&01/07&02/07&03/07&04/07\\ \hline
\rowcolor{LGray} $\text{train}_u$ & 2121 & 2121& 2121& 2121& 2121& 2121& 2121& 2121& 2121\\
 $\text{train}_t$ & 307K & 395K & 468K & 543K & 609K & 685K & 752K & 814K & 867K \\
\rowcolor{LGray} $\text{test}_u$  & 1804 & 1985 & 2045 & 2115 & 2146 & 2174 & 2184 & 2194 & 2197 \\
 $\text{test}_t$ & 293K & 358K & 414K & 477K & 533K&599K & 658K & 718K &  768K   \\\hline
\end{tabular}
}
\end{table}

% 
% Setting $n=0$, would imply that we are assigning the label ``YES'' if the user has re-tweeted more YES-supporting accounts (and inversely), which might result into a low quality training set, whereas higher values for $n$ would imply a smaller (but of higher quality) training set.

\subsection{Test Set}
\label{sec:52}

\iffalse
\begin{table}[!t]
\centering
\resizebox{\columnwidth}{!}{
\begin{tabular}{|l||c|l|rr|c|}\hline
    & \begin{tabular}[c]{@{}c@{}}Annotated\\ users\end{tabular}&\begin{tabular}[c]{@{}c@{}}Agreement\\ users (tweets)\end{tabular}&YES & NO &  $\kappa$ \\ \hline
\#1 &2,700 &2,365 (1.3M)&514&1,683 &   .75            \\
%\#2 &3,718 &3,216 (??)&369&938     &   .76    \\
\hline
\end{tabular}}
\caption{My caption}
\label{my-label}
\end{table}
\fi

%in an attempt to monitor the voting intention of \textit{(a)} highly active and vocal users who are driving the political conversation on Twitter and \textit{(b)} a more representative (random) sample of Twitter users. 

% during the Periods 2--4

For evaluation purposes, we generate a test set of active users that are likely to participate in political conversations on Twitter. First, we identify all users having tweeted at least 10 times after the referendum announcement (86,000 users). From the 500 most popular hashtags in their tweets, we selected those that were clearly related to the referendum (189) which were then manually annotated with respect to potentially conveying the user's voting intention (e.g., ``yesgreece'', ``no'' as opposed to neutral ones, such as ``referendum''). Finally, we selected a random sample of 2,700 users (out of 22K) that had used more than three such hashtags, to be manually annotated -- without considering any user from the training set. This is standard practice in related work~\cite{Garimella2018,Stewart2018} and enables us to evaluate our models on a high quality test set, as opposed to previous related work which rely on keyword-matching approaches to generate their test set \cite{fang2015topic,zubiaga2017stance}. 
 
Two authors of the paper (Greek native speakers) annotated each of the users in the test set, using the tweets after the referendum announcement. Each annotator was allowed to label an account as \texttt{YES}, \texttt{NO}, or \texttt{N/A}, if uncertain. There was an agreement on 2,365 users (Cohen's $\kappa=.75$) that is substantially higher if the \texttt{N/A} labels are not considered ($\kappa=.98$), revealing high quality in the annotations, i.e., in the upper part of the `substantial' agreement band~\cite{artstein2008inter}. We discarded all accounts labelled as \texttt{N/A} by an annotator and used the remaining accounts where the annotators agreed for the final test set, resulting to 2,197 users -- similar test set sizes  are used in related tasks \cite{derczynski2017semeval}. The resulting user distribution (\texttt{NO} 77\%, \texttt{YES} 23\%) is more imbalanced compared to the actual result of the referendum, due to the demographic bias on Twitter \cite{miranda2015twitter}. To mimic a real-time scenario, we refrained from balancing our train/test sets, since it would have been rather impossible to know the voting intention distribution of  Twitter users a-priori. Overall, we use 18.9\% (1.64M/8.66M) of the tweets written in Greek during that period in our experiments (see Table \ref{tab:testusers}).%before the announcement of the referendum.

\section{Models}
\label{sec:methods}

\subsection{Convolution Kernels}

Convolution kernels are composed of sub-kernels operating on the item-level to build an overall kernel for the object-level \cite{haussler1999convolution,collins2002convolution} and can be used with any kernel based model such as Support Vector Machines (SVMs)~\cite{Joachims1998}. Such kernels have been applied in various NLP tasks \cite{collins2002convolution,kim2015convolutional,tymoshenko2016convolutional,lukasik2016convolution}. %\citet{lukasik2016convolution} employed a similar approach for streaming online documents. 
Here we build upon the approach of \citet{lukasik2016convolution} by combining convolution kernels operating on available (1) \emph{text}; and (2) \emph{network information}. 

% Let $a$, $b$ denote two users in a social network, posting messages $W_a=\{w_{a}^{1},...,w_{a}^{N}\}$ and $W_b=\{w_{b}^{1}$, ..., $w_{b}^{M}\}$ with associated timestamps $T_a=\{t_{a}^{1},...,t_{a}^{N}\}$ and $T_b=\{t_{b}^{1}$, ..., $t_{b}^{M}\}$ respectively. A message $w_{i}^{j}$ of user $i$ at time $j$ can be represented by any $k$-dimensional dense vector (see Section~\ref{sec:experiments}).

Let $a$, $b$ denote two objects (e.g., social network users), represented by two $M\times N$ matrices $Z_a$ and $Z_b$ respectively, where $M$ denotes the number of items representing the object and $N$ the dimensionality of an item vector. For example, an item can be a user's tweet or network information. A kernel $K$ between the two objects (users) $a$ and $b$ over $Z_a$ and $Z_b$ is defined as:
\begin{equation}\label{eq:1}
    K_z(a, b) = \frac{1}{\left|Z_a\right|\left|Z_b\right|}\sum_{i, j} k_z(z_{a}^{i}, z_{b}^{j}),
\end{equation}
where $k_z$ is any standard kernel function such as a linear or a radial basis function (RBF). One can also normalise $K_z$ by dividing its entries $K_z(i, j)$ by $\sqrt{K_z(i, i) K_z(j, j)}$.

The resulting kernel has the ability to capture the similarities across objects on a per-item basis. However, unless restricted to operate on consecutive items (time-wise), it ignores their temporal aspect. Given a set of associated timestamps $T_o=\{t_{o}^{1},...,t_{o}^{N}\}$ for the items of each object $o$, \citet{lukasik2016convolution} proposed to combine the temporal and the item aspects as:
\begin{equation}\label{eq:2}
  K_{zt}(a, b)=\frac{1}{\left|Z_a\right|\left|Z_b\right|}\sum_{i, j}k_z(z_{a}^{i}, z_{b}^{j})k_t(t_{a}^{i}, t_{b}^{j}),
\end{equation}
where $k_{t}$ is any valid kernel function operating on the timestamps of the items. Here, $K_{zt}$ is a matrix capturing the similarities across users by leveraging both the information between pairs of items and their temporal interaction.

\subsubsection{Text Kernels} 

Let $a$, $b$ denote two users in a social network, posting messages $W_a=\{w_{a}^{1},...,w_{a}^{N}\}$ and $W_b=\{w_{b}^{1}$, ..., $w_{b}^{M}\}$ with associated timestamps $T_a=\{t_{a}^{1},...,t_{a}^{N}\}$ and $T_b=\{t_{b}^{1}$, ..., $t_{b}^{M}\}$ respectively. We assume that a message $w_{i}^{j}$ of user $i$ at time $j$ is represented by the mean $k$-dimensional embedding~\cite{mikolov2013distributed} of its constituent terms. %(for more details refer to Section~\ref{sec:experiments}). 
This way, we can obtain \emph{text convolution kernels}, $K_{w}$ and $K_{wt}$ by simply replacing $Z$ and $z$ with $W$ and $w$ respectively in Equations \ref{eq:1} and \ref{eq:2}. Following \citet{lukasik2016convolution}, we opted for a linear kernel operating on text and an RBF on time.

\subsubsection{Network Kernels} 

Let assume a set of directed weighted graphs $\textstyle G=\{G_1(N_1, E_1), ..., G_t(N_t, E_t)\}$, where $G_i(N_i, E_i)$ represents the retweeting activity graph of the $N_i$ users at a time point $ i\in T=\{1, .., t\}$. Let $L_a \in \mathbf{R}^{N,k}$, $L_b \in \mathbf{R}^{M,k}$ denote the resulting matrices of a k-dimensional, network-based user representation for two users $a$ and $b$ across time. Contrary to the textual vector representation $w_{i}^{j}$ that is defined over a fixed space given a pre-defined vocabulary, user network vector representations (e.g., graph embeddings \cite{tang2015line}), are computed at each time step on a different network structure. Thus, a standard similarity score between two user representations at timepoints $t$ and $t$+1 cannot be used, since the network vector spaces are different. To accommodate this, at each time point $t$ we calculate the median $L_{\texttt{YES}}^t$ and $L_{\texttt{NO}}^t$ vectors for each class of our training examples and update the respective user vectors as: 
\begin{align*}
    \textstyle L_u^{*t} = d(L_{\texttt{YES}}^t, L_u^t) - d(L_{\texttt{NO}}^t, L_u^t),
\end{align*}
%$L_u^{*t}=d(L_{\texttt{YES}}^t, L_u^t)-d(L_{\texttt{NO}}^t, L_u^t)$, 
using some distance metric $d$  (for simplicity, we opted for the linear distance). If a user has not retweeted, his/her original network representation $l_u^{t}$ is calculated as the average across all user representations at $t$. Finally, the \emph{network convolution kernels}, $K_n$ and $K_{nt}$ are computed using Equations \ref{eq:1} and \ref{eq:2} respectively by simply replacing $Z$ with $L^*$ and $z$ with $l^*$. Similarly to text kernels, we use a linear kernel $k_n$ for the network and an RBF kernel $k_t$ for time.

\subsubsection{Kernel Summation}

We can combine the text and network convolution kernels by summing them up: $K_{sum} = K_w+K_{wt}+K_n+K_{nt}$. This implies a simplistic assumption that the contribution of the different information sources with respect to our target is equal. While this might hold for a small number of carefully designed kernels, it lacks the ability to generalise over multiple kernels of potentially noisy representations.

\subsection{Convolution Kernel Models}
\subsubsection{SVMs with Convolution Kernels} 
Convolution kernels can be used with any kernel based model. Here, we use them with SVMs. First, a SVM$_{s}$ operates on a single information source $s=\{w,n\}$, i.e., SVM$_{w}$ for text and SVM$_{n}$ for network. Second, a SVM$_{st}$ takes temporal information into account combined with text (SVM$_{wt}$) and network (SVM$_{nt}$) information respectively. Finally, we combine the text and the network information using a linear kernel summation ($K_{sum}$) of their respective kernels (SVM$_{sum}$).

\subsubsection{Multiple Convolution Kernel Learning (MCKL)} 

Multiple kernel learning methods learn a weight for each kernel instead of assigning equal importance to all of them allowing more flexibility.
%learn a model alongside the weights of its sub-kernels, overcoming the danger of feeding a kernel-based model with a na\"ive summation of potentially noisy kernels. 
Such approaches have been extensively used in tasks where different data modalities exist \cite{jaques2015multi,tsakalidis2016combining,poria2017ensemble}. We build upon the approach of \citet{sonnenburg2006large} to build a model based on labelled instances $x_i \in I$, by combining the different convolution kernels $K_s$ with some weight $w_s>0$ s.t. $\sum_{s}{w_s}$=1 and apply: 
\begin{align*}
 f(x)=sign\bigg(\sum_{i\in I}\alpha_i\sum_{s} w_s K_s(x,x_i) + b\bigg).   
\end{align*}
The parameters $\alpha_i$, the bias term $b$ and the kernel weights are estimated by minimising the expression:
%($\gamma-\sum_{i}\alpha_i$), with respect to $\gamma\in R$, $\alpha\in R_+^{|I|} \nonumber$, such that $0\leq\alpha_i\leq C \;\forall i$, $\sum_{i\in{I}} \alpha_i y_i=0 \nonumber$, $\frac{1}{2}\sum_{i\in{I}}\sum_{j \in{I}} \alpha_i \alpha_j y_i y_j K_s({\bf x}_i,{\bf x}_j)\leq \gamma\;\; \forall s$. 

\begin{eqnarray*} 
%\hspace{-1cm}
\mbox{min} && \gamma-\sum_{i\in{I}}\alpha_i\\ \mbox{w.r.t.} && \gamma\in R, \alpha\in R_+^{|I|} \nonumber\\ \mbox{s.t.} && 0\leq\alpha_i\leq C \;\forall i,\;\;\sum_{i\in{I}} \alpha_i y_i=0 \nonumber\\ && \frac{1}{2}\sum_{i\in{I}}\sum_{j \in{I}} \alpha_i \alpha_j y_i y_j K_s({\bf x}_i,{\bf x}_j)\leq \gamma\;\; \forall s.\\ \end{eqnarray*}

This way, the four convolution kernels are calculated individually and subsequently combined in a weighted scheme accounting for their contribution in the prediction task. This allows us to combine external and asynchronous information (e.g., news articles), while adding other kernels capturing different aspects of the users (e.g., images) is straight-forward.

\section{Experimental Setup}
\label{sec:experiments}

%We extract features derived from the tweets (``TEXT'') and the re-tweeting activity (``NETWORK'') of the users in our training and test set. In our experiments, we use each of these two sources as input to various models individually, as well as in concatenation (``BOTH''). 

\subsection{Features}

\subsubsection{Textual Information (TEXT)} 

We obtain word embeddings by training \texttt{word2vec} \cite{mikolov2013distributed} on a collection of 14.7 non-retweeted tweets obtained by \cite{tsakalidis2018building}, collected in the exact same way as our dataset, over a separate time period. We performed standard pre-processing steps including lowercasing, tokenising, removal of non-alphabetic characters, replacement of URLs, mentions and all-upper-case words with identifiers. We used the CBOW architecture, opting for a 5-token window around the target word, discarding all words appearing less than 5 times and using negative sampling  with 5 ``noisy'' examples. After training, each word is represented as a 50-dimensional vector. Each tweet in our training and test set is represented by averaging each dimension of its constituent words.

%Feature extraction on the tweet level is performed by averaging each dimension of its words. For our baselines, we extract the average embedding values across all tweets of a user up to the current evaluation point.

\subsubsection{Network Information (NETWORK)} 

We trained LINE \cite{tang2015line} embeddings at different timesteps, by training on the graphs $\{G_1(N_1, E_1), ..., G_T(N_T, E_T)\}$, where $N_i$ is the set of users and $E_i$ is the (directed, weighted) set of retweets amongst $N_i$ up to time $i$. We choose the ``retweet'' rather than the ``user mention'' network
%since retweets are more likely to be endorsements
due to its more polarised nature, as indicated by past work \cite{Conover2011}\footnote{The ``following'' network cannot be constructed based on the JSON objects returned by Twitter Streaming API; to achieve this requires a very large number of API calls and cannot be constructed accurately in a realistic scenario.}. LINE was preferred over alternative models \cite{perozzi2014deepwalk,rizos2017multilabel} due to its ability to model directed weighted graphs. We construct the network $G_t$ every 12 hours based on the retweets among all users up to time $t$,
%. An alternative approach would have been to construct the network in a sliding window approach; however, finding the optimal value for its width can be a crucial context-dependent task, which opposes our goal of generalisation ability. 
and LINE is trained on $G_t$ to create 50-dimensional user representations.
%adtsakal:for-appendix supplementary material
We used the second-order proximity, since it performed better than the first-order in early experimentation. We also refrained from concatenating them to keep the dimensionality relatively low. 

%Our baselines employ only the user embeddings at the current evaluation time point; our models take all of the past representations into account to build the  $K_n$, $K_{nt}$ kernels.

\subsection{Models}
\subsubsection{Convolution Kernel Models}

Our MCKL and our SVM models are fed with the convolution kernels operating on the tweet-level (for TEXT) and each NETWORK representation (derived every 12 hours), based on the tweets and re-tweeting activity respectively of the users up to the current evaluation time point. %If a user has not participated in the re-tweet network up to the current evaluation point in time $t$, his/her $L_u$ representation at time $t$ is calculated as the average user representation at $t$ -- if he/she has not tweeted until $t$, then he/she is ignored for the evaluation at $t$.

\subsubsection{Baselines} 

We compare our proposed methods against competitive baselines that are commonly used in social media mining tasks trained on feature aggregates \cite{ma2015detect,zubiaga2017stance}. We obtain a TEXT representation of a user at each time step $t$ by averaging embedding values across all his/her tweets until $t$. Similarly, a user NETWORK representation is computed from the retweeting graph up until $t$. Finally, we train a regularised Logistic Regression ({\bf LR}) with $L_2$ regularisation~\cite{Le1992}, a feed-forward neural network ({\bf FF})~\cite{Hornik1989}, a Random Forest ({\bf RF})~\cite{Breiman2001} and a {\bf SVM}.

%For every evaluation time step $t$, we merge all the tweets by every user and form a network based on their retweets until $t$. We perform feature extraction (see below) and apply Logistic Regression (LR), Random Forest (RF) and Support Vector Machine (SVM). 

%adtsakal:for-appendix
%
\paragraph{Model Parameters} 

Parameter selection of our models and the baselines is performed using a 5-fold cross-validation on the training set. We experiment with different regularisation strength ($10^{-3}, 10^{-2}, $ $..., 10^{3}$) for LR, different number of trees (50, 100, ..., 500) for RF, and different kernels (linear, RBF) and parameters C and $\gamma$ ($10^{-3}, 10^{-2}, ..., 10^{3}$) for SVMs. For FF, we stack dense layers, each followed by a ReLU activation and a 20\% dropout layer, and a final layer with a sigmoid activation function. We train our network using the Adam optimiser \cite{kingma2014adam} with the binary cross-entropy loss function
and experiment with different number of hidden layers (1, 2), units per layer (10, 25, 50, 75, 100, 150, 200), batch size (10, 25, 50, 75, 100) and number of epochs (10, 25, 50, 100). For MCKL, we experiment with the same C values as in SVM and apply an $L_2$ regulariser.

\begin{figure*}[!t]
  \centering
  \subfloat{\includegraphics[scale=0.41]{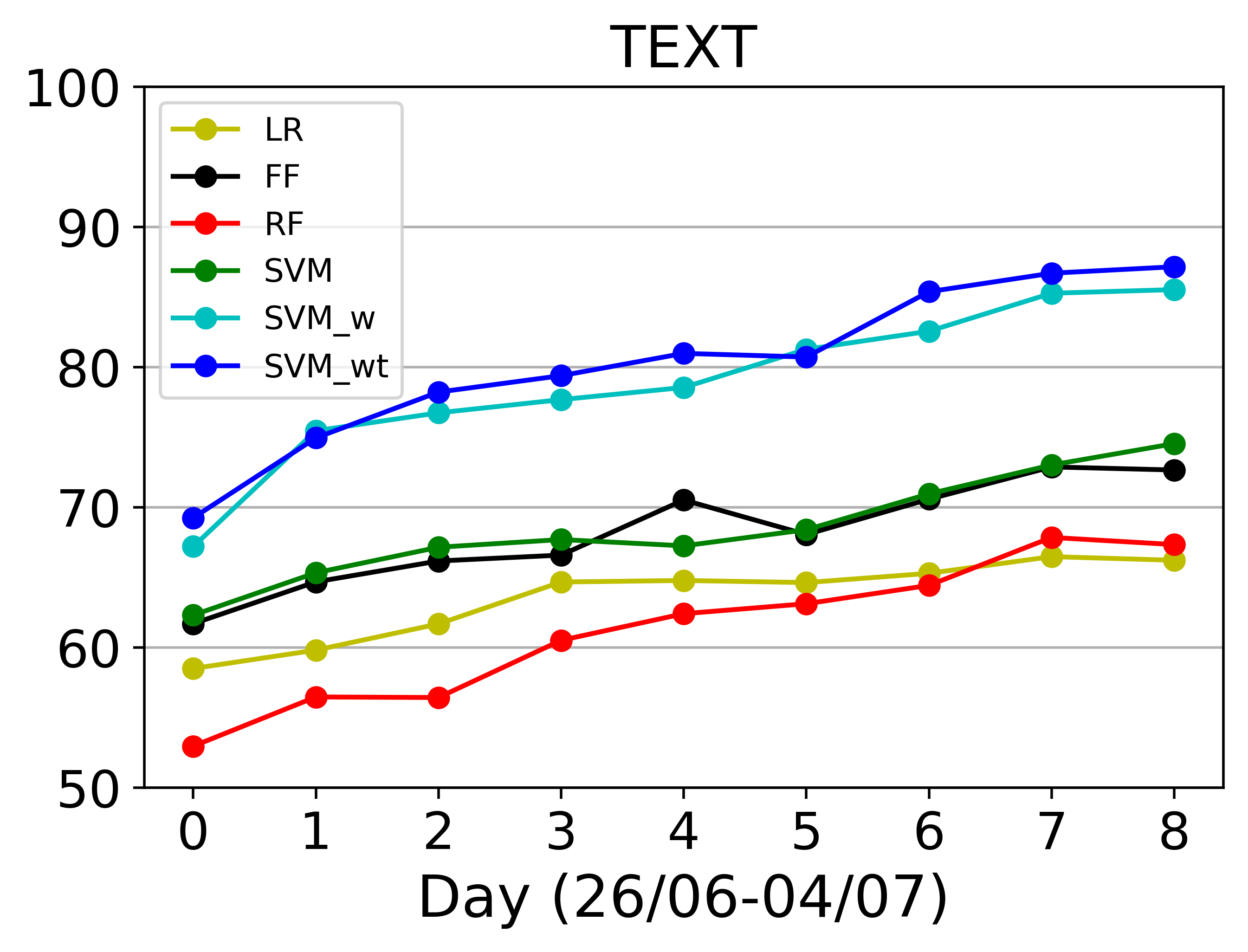}}\quad %315
  \subfloat{\includegraphics[scale=0.41]{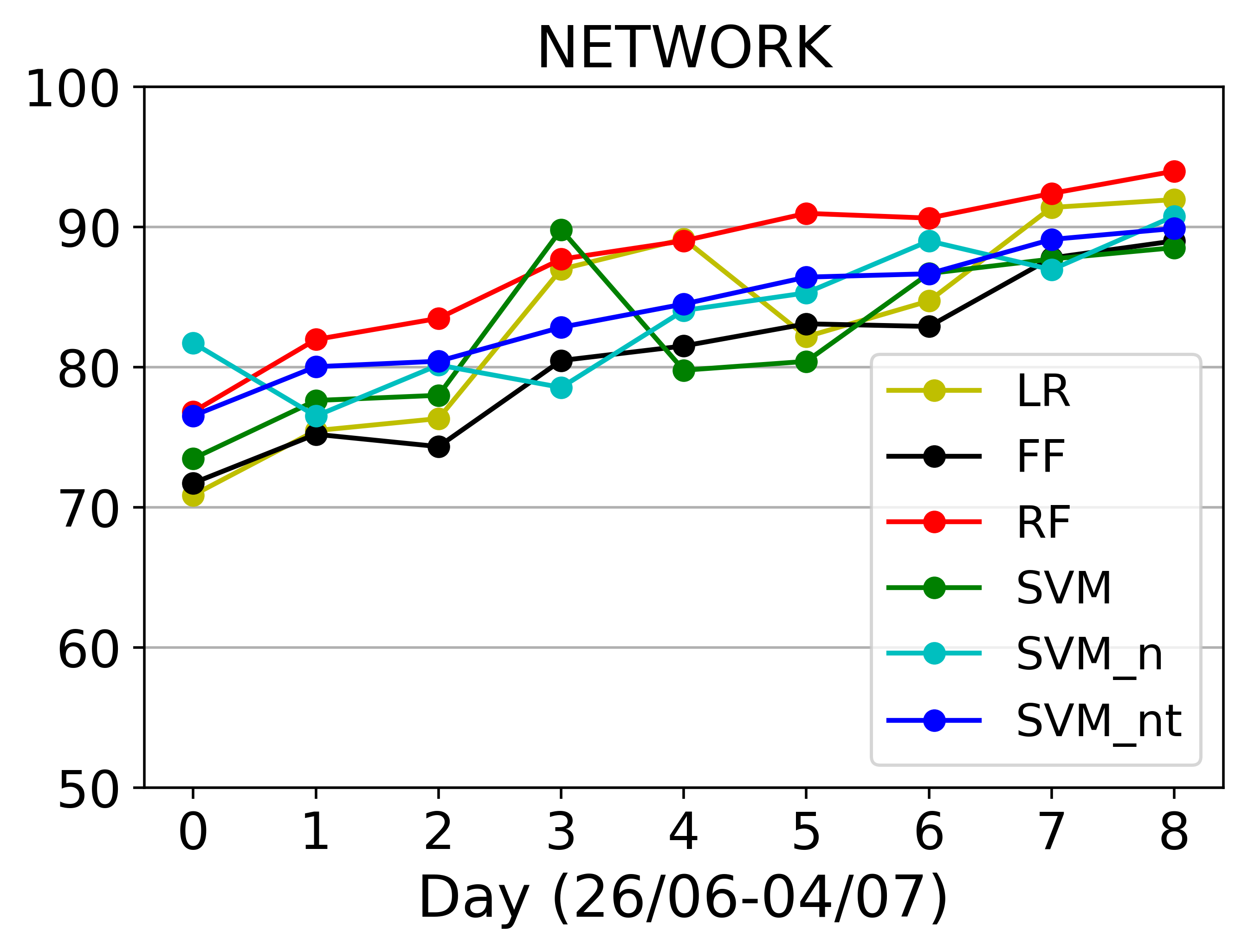}}\quad
  \subfloat{\includegraphics[scale=0.41]{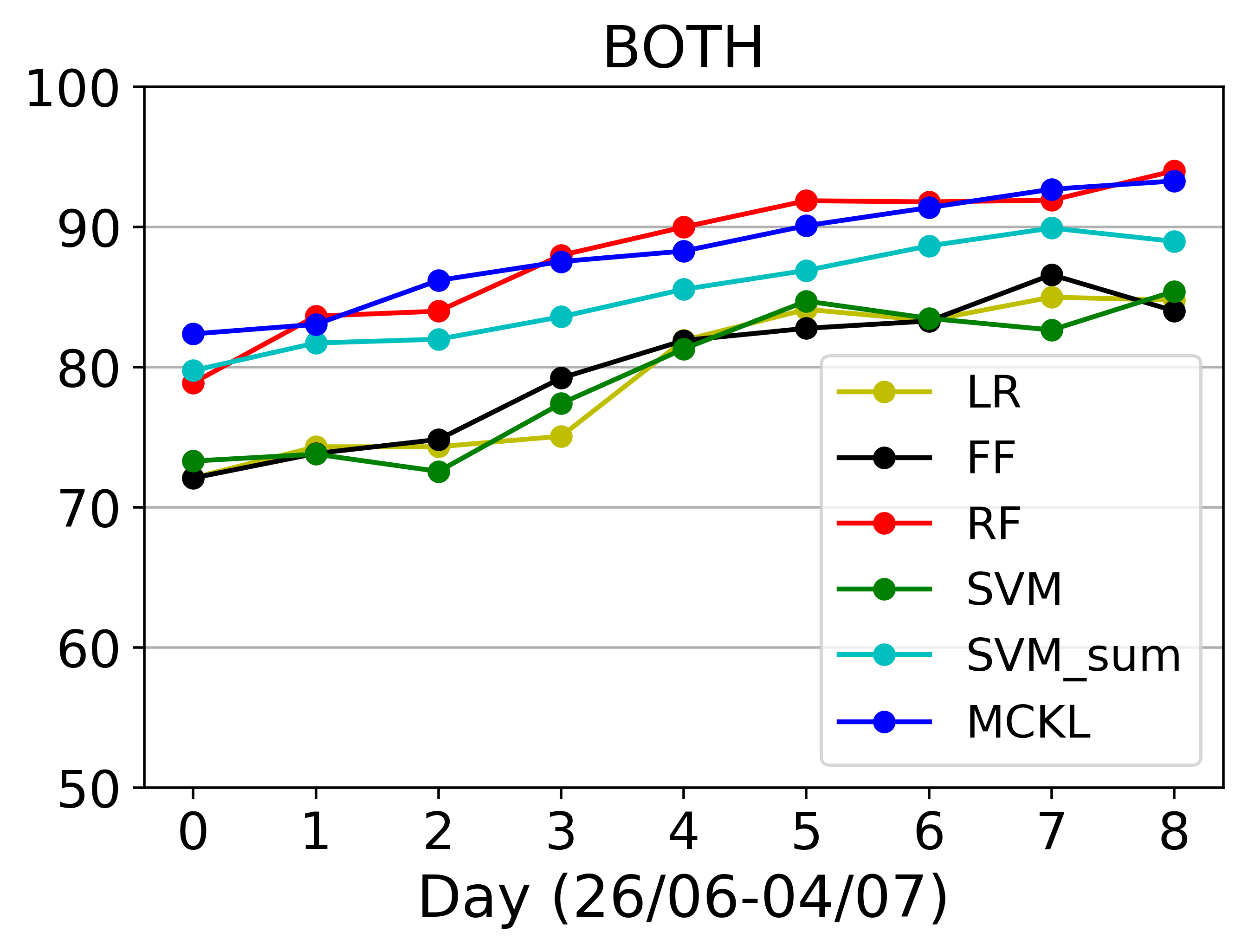}}\quad
\caption{Macro-average F-score across all evaluation days using \textbf{TEXT}, \textbf{NETWORK} and \textbf{BOTH} user representations.}
\label{fig:results}
\end{figure*}

\subsection{Evaluation} 

We train and test our models based on the data collected on a daily basis (every midnight), starting from the day before the announcement of the referendum (day 0) until the day before its due date (day 8). This way, we mimic a real-time setting and gain better evaluation insights. To evaluate our models, we compute the macro-average F-score, which forms a more challenging metric compared to micro-averaging, given the imbalanced distribution of our test set. At each evaluation time point $t$, we use information about the users in our training set up to $t$, to classify the test users that have tweeted at least once up to $t$ (note that all of the users in our training set have tweeted before the announcement of the referendum, thus the size of the training set in terms of number of users remains constant). This results into a different number of test instances per day (see Table~\ref{tab:testusers}). However, we did not observe any major differences in our evaluation by excluding newly added users. Parameter selection is performed on every evaluation day using a 5-fold cross-validation on the training set.

\section{Results}

\subsection{Nowcasting Voting Intention} 

Figure~\ref{fig:results} presents the macro-average F-scores obtained by the methods compared in all days from the announcement to the day of the referendum. As expected, the closer the evaluation is to the referendum date, the more accurate the models since more information becomes available for each user. Table~\ref{tab:newtable} shows the average (across-all-days) F-score by each model.

\begin{table}[t]
\renewcommand{\arraystretch}{.9}
\centering
    \caption{Average F-score and standard deviation across all evaluation days using \textbf{TEXT}, \textbf{NETWORK} and \textbf{BOTH} user representations. SVM$_{s}$ and SVM$_{st}$ denote the SVM with convolution kernels (SVM$_{w}$, SVM$_{n}$) and (SVM$_{wt}$, SVM$_{nt}$), respectively.}
    \label{tab:newtable}
\resizebox{\columnwidth}{!}{
    \begin{tabular}[b]{lccc}
      \hline
      & \textbf{TEXT} & \textbf{NETWORK} & \textbf{BOTH} \\ \hline 
      LR  & 63.55 $\pm$2.86& 83.21$\pm7.55$ & 79.43$\pm5.34$\\
      FF  & 68.19 $\pm3.78$& 80.66$\pm5.93$ & 79.83$\pm5.11$\\
      RF  & 61.27 $\pm5.14$& \textbf{87.43$\pm5.60$} & 88.22$\pm5.03$\\
      SVM & 68.51 $\pm3.80$&82.43$\pm5.83$ & 79.39$\pm5.18$\\ \hline
      SVM$_s$  & 78.91$\pm5.68$ & 83.65$\pm4.82$ & --\\
      SVM$_{st}$  & \textbf{80.30$\pm5.81$} & 84.03$\pm4.47$ & -- \\ 
      SVM$_{sum}$ & -- & -- & 85.22$\pm3.64$\\
      MCKL & -- & -- & \textbf{88.31$\pm3.95$} \\ \hline
    \end{tabular}}

\end{table}

Temporal convolution kernels using \textbf{TEXT} (SVM$_{wt}$) significantly outperform the best text-based baseline ($p=.001$, Kruskal-Wallis test against SVM), with an average of 11.8\% and 17.2\% absolute and relative improvement respectively. This demonstrates the model's ability on capturing the similarities between different users on a per-tweet basis compared to simpler models using tweet aggregates. Also, SVM$_w$ and SVM$_{wt}$ implicitly capture similarities in the retweeting activity of the users. This is important, since network information might not be easily accesible (e.g., due to  API limitations) while it is expensive to compute at each timestep. Hence, one can use SVM$_{wt}$ to model user written content and partially capture network information. %(i.e. retweets are included in \textbf{TEXT} representation).

Classification accuracy consistently improves when using the \textbf{NETWORK} representation (i.e., graph embeddings). RF achieves 94\% F-score on the day before the referendum, whereas the worst-performing baseline (FF) still achieves 80.66\% F-score on average. SVM$_{nt}$ provides a small boost (1.6\% on average) compared to the vanilla SVM, which uses only the user representations derived at the current time point. This implies that the current network structure is indicative of users' voting intention, probably because the referendum was the dominant topic of discussion at the time, e.g., most of the retweeting activity was relevant the referendum (see Section~\ref{sec:analysis}). This is also in line with recent findings of \citet{Aletras2018} on predicting occupation class and income where network information is more predictive than language.

When combining the user text and network representation (\textbf{BOTH}), the baselines fail to improve over using only NETWORK. In contrast, our MCKL improves by 4.28\% over the best performing single convolution kernel model (SVM$_{nt}$). This demonstrates that MCKL can effectively combine information from both representations by weighting their importance, and further improve the accuracy of the best performing single representation model. Overall, MCKL significantly outperforms the best performing text-based baseline by approximately 20\% in F-score ($p<.001$, Kruskal-Wallis test).

\subsection{Robustness Analysis} 

\begin{figure}[!t]
 \centering
 \includegraphics[width=.72\columnwidth]{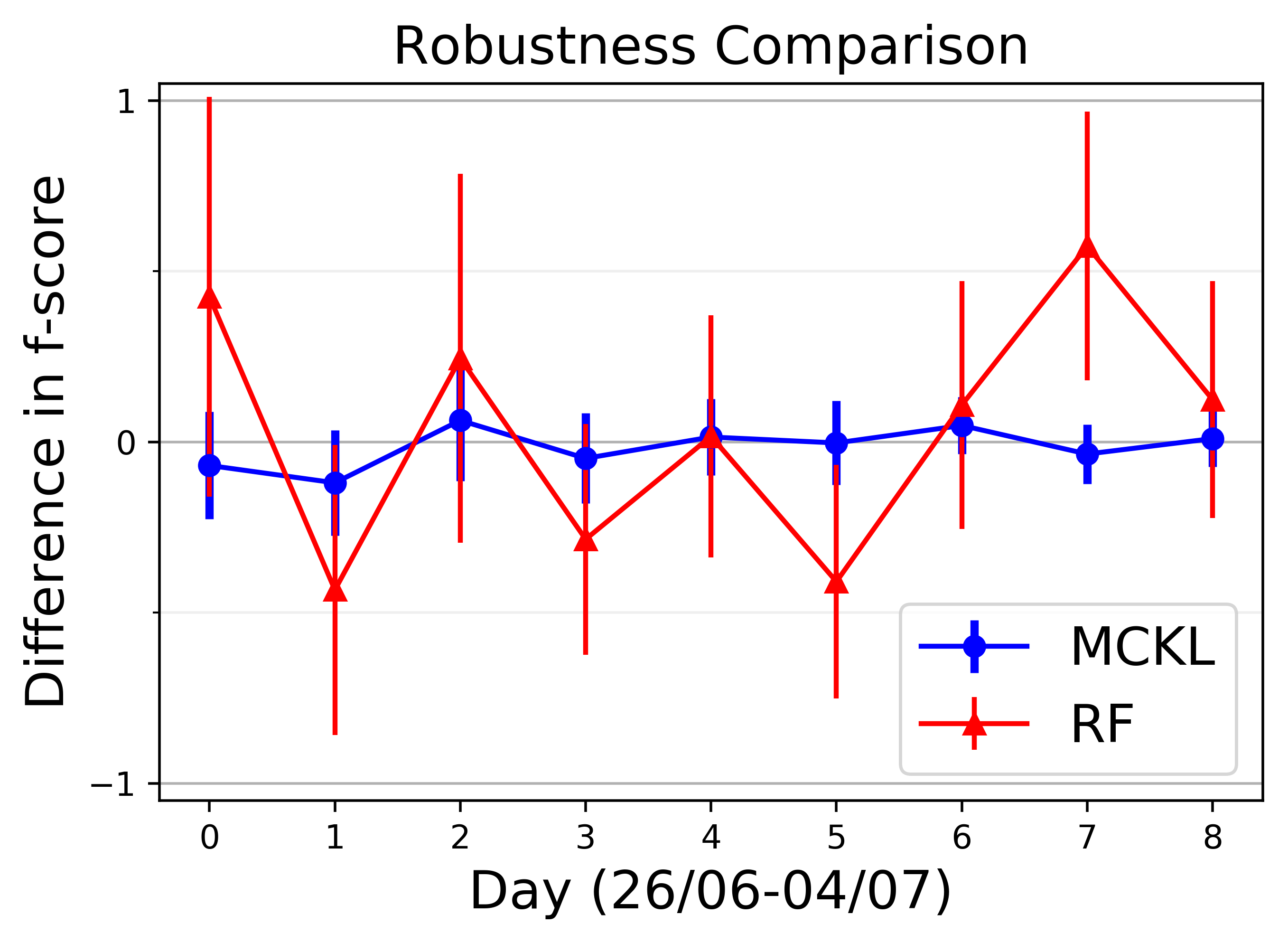}
   \caption{Change in performance (mean/standard deviation) compared to the results in Figure~\ref{fig:results}, after 100 experiments with added noisy features.}
 \label{fig:robust}
 \end{figure}

Due to the semi-supervised nature of our task, it is impossible to judge whether the small difference between MCKL and RF stems from a better designed model. Furthermore, it is difficult to assess MCKL's effectiveness with respect to its ability to generalise over multiple and potentially noisy feature sources. 

To assess the robustness of the best performing models (MCKL, RF) operating on \textbf{BOTH} information sources, here we perform experiments by adding random noise in their input. We assume that there is a noisy source generating an extra K-dimensional representation $X$ for every user that we add as extra input to the models. We set $K=25$, so that \textit{(a)} we account for a smaller noisy input compared to our features (25 vs 50) and \textit{(b)} 1/5 of our kernels in MCKL and 25/125 input features in RF are noisy. We perform 100 runs, each time drawing random noise $X\sim N(0, 1)$. 

Our results indicate that RF is more sensitive to the noisy input compared to MCKL (see Figure~\ref{fig:robust}). On average, RF achieves a small boost (0.04\%) in performance with the added noise. That together with the higher standard deviation reveal the vulnerability of RF to potentially corruption and stochasticity introduced in the input. On the contrary, MCKL is consistently robust, achieving only a tiny reduction in performance on average across all days (0.02\%) while the respective average standard deviation is lower than the one achieved by RF (0.12 vs 0.41). This robustness is highly desirable is cases of such sudden political events and it also indicates that we can add kernels capturing different properties of our task (e.g., user-related information, images, etc.), without having to decide a-priori which of them are indeed predictive of the user's voting intention. We plan to investigate this in future work.

\section{Qualitative Analysis}
\label{sec:analysis}

In this section, we provide insights into the temporal variation observed in the users' shared content and the network structure during this major political crisis. Besides performing a qualitative analysis during this time period, we believe that this analysis will also provide insights on \textit{(a)} the reasons that trigger the significant improvement in performance of convolution kernels methods operating on TEXT, and \textit{(b)} the reason that our non-temporally-sensitive baselines are rather competitive to our convolution kernel models, when using NETWORK information. In the current section we provide details on both of these aspects.

\subsection{Language} 

We are interested in investigating which are the political-related entities  that voters from both sides most likely mention. We expect that this will shed light on 
%whether 
the main focus of discussion in the political debates between the \texttt{YES}/\texttt{NO} voters that occurred after the announcement of the referendum.
% was the opposite side or not. 
For this, two authors manually compiled two lists of n-grams containing different ways of referring\footnote{Note that Greek is a fully inflected language. We opted not to apply stemming because inflected word forms carry meaningful information.} to the \textit{(a)} the six major political parties and \textit{(b)} their leaders (see Table~\ref{tab:parties}). 
We represent every \texttt{YES}/\texttt{NO} user in the test set as aggregated \texttt{tf-idf} values of the ngrams (1-3) appearing in his/her concatenated tweets; then, we compute an n-gram 's $n$ score as $PMI(\textit{n}, \texttt{YES}) - PMI(\textit{n}, \texttt{NO})$. A positive score implies that it is highly associated with users who support the \texttt{YES} vote, and vice versa.

\begin{figure}[!t]
\centering
\includegraphics[width=.94\columnwidth]{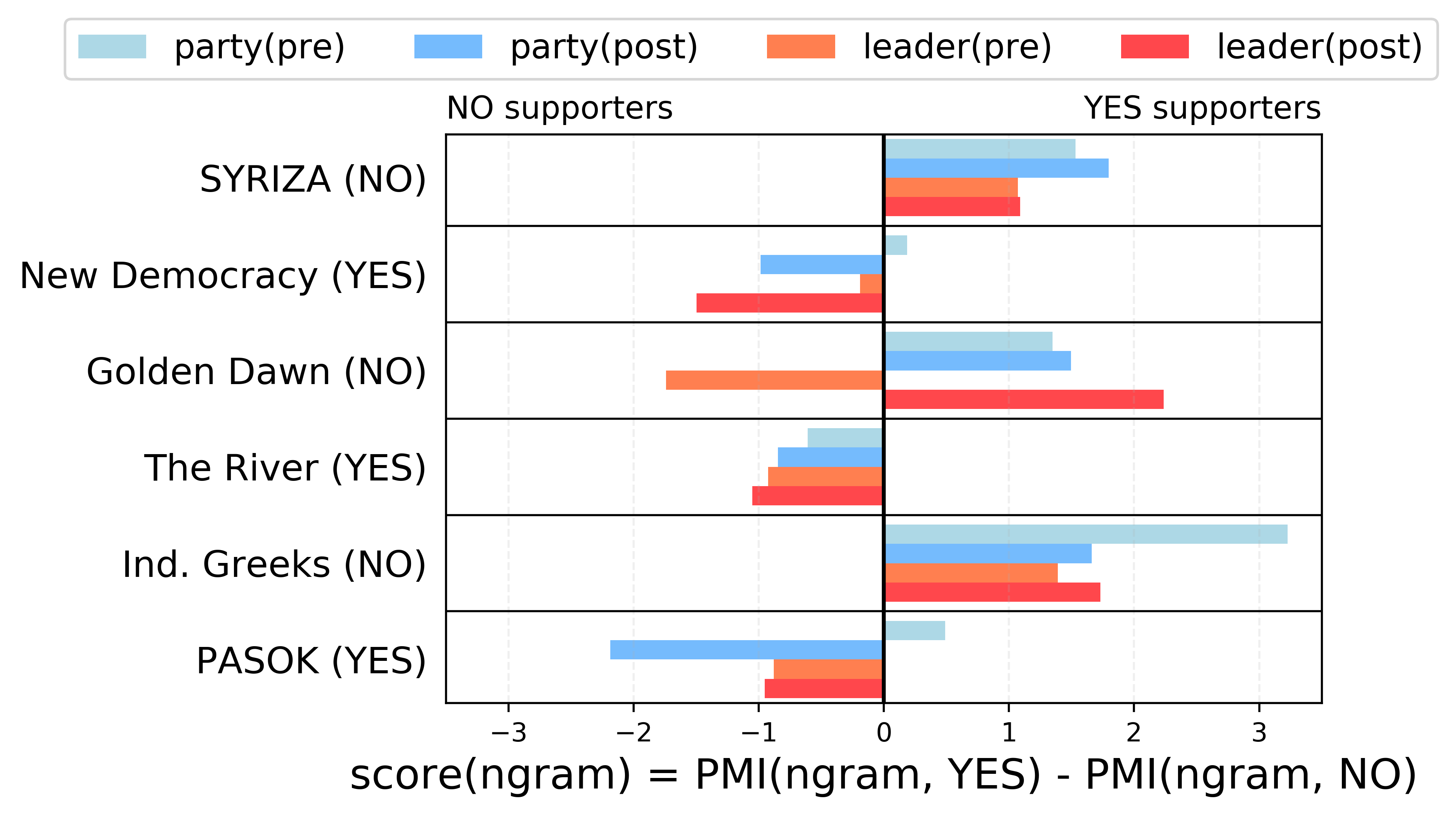}
\caption{Scores of n-grams related to the political parties/leaders, pre (18/06-26/06) and post (27/06-05/07) the referendum announcement. %Scores$<$0 ($>$0) indicate that n-grams appear mostly  in tweets of \texttt{NO} (\texttt{YES}) voters.}
}
\label{fig:text_discuss}
\end{figure}

Figure~\ref{fig:text_discuss} shows that the parties and leaders that supported one side, mostly appear in tweets of users supporting the opposite side. This is more evident when we consider tweets shared by the users \textit{after} the announcement of the referendum. Examining the content of highly-retweeted tweets, revealed sarcasm and hostility for the opposite side in the majority of them (see Table~\ref{tab:sarcastic}). Hostility is a frequent phenomenon in public debates~\cite{Jorgensen1998} and our findings corroborate previous work showing that the political discourse on Twitter is polarised~\cite{Conover2011,Garimella2018}.

\begin{table}[!t]
\centering
\caption{Most similar words to \texttt{YES} and \texttt{NO} (translated to English), when training word2vec on  different time periods.}
\label{tab:language}
\renewcommand{\arraystretch}{1.2}
\resizebox{\columnwidth}{!}{\begin{tabular}{lcc}
    &{\bf Before the announcement} &{\bf After the announcement}\\ 
    & (18/06-26/06) & (27/06-05/07)\\ \hline
{\bf YES} & \begin{tabular}[c]{@{}l@{}}no, ok, nah, alright, sure,\\usrmnt, hahaha, alrighty,\\but, so\vspace{.2cm}\end{tabular}                                  & \begin{tabular}[c]{@{}l@{}}no, abstain, referendum, KKE,\\question, invalid, euro, clearly,\\clash, nai\vspace{.2cm}\end{tabular} \\ 
\hline
{\bf NO}  & \begin{tabular}[c]{@{}l@{}}yes, only, sure, so (slang),\\disagree, mainly, especially,\\obviously, so (abbrv), agree\end{tabular} & \begin{tabular}[c]{@{}l@{}}yes, abstain, KKE, referendum,\\ clash, question, people, invalid,\\vote, clearly\end{tabular}\\
\end{tabular}}

\end{table}

\begin{figure}[!t]
 \centering
 \includegraphics[width=.9\columnwidth]{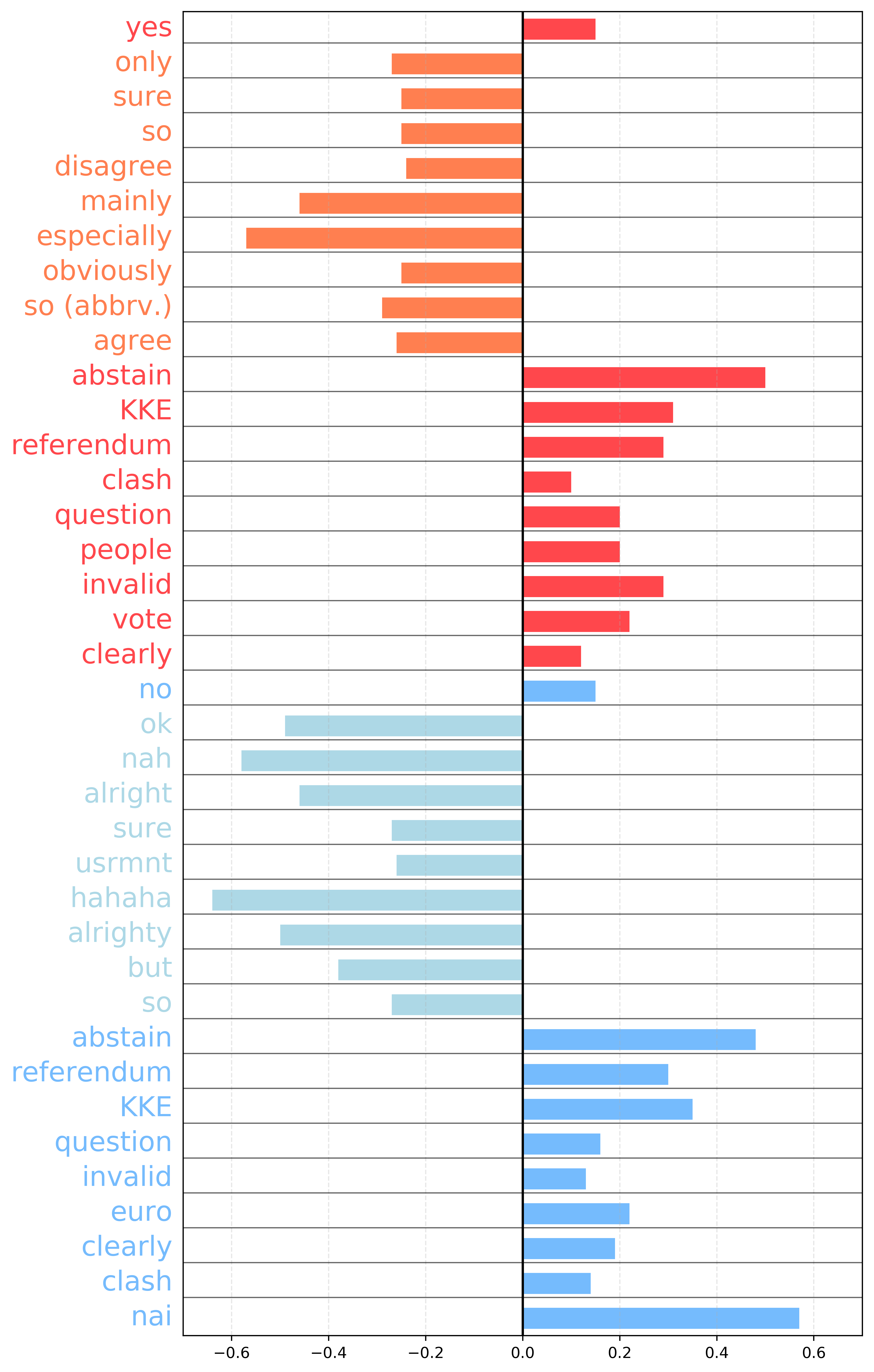}
 \caption{Difference in cosine similarity ($cos_{post} (w_{no/yes}, w) - cos_{pre} (w_{no/yes}, w)$) between the \emph{no}/\emph{yes} (red/blue) word vectors $w_{no/yes}$ and each of their most similar words in the two periods.}
 \label{fig:words_before_after}
 \end{figure}

Finally, we examine the temporal variation of language over the same two periods. Table~\ref{tab:language} shows the most similar words (translated to English) to the \emph{yes} and \emph{no} words, measured by cosine similarity, when training \texttt{word2vec} using the tweets of each time period. The difference of the cosine similarities $cos_{post} - cos_{pre}$ between the \emph{yes}/\emph{no} vectors and each of their corresponding most similar words over these two periods is shown in Figure~\ref{fig:words_before_after}. After the announcement, the context of the two words shifts towards the political domain. That might explain why text aggregates become noisy, as shown in our results. Convolution kernels are able to filter-out this noise since they operate on the tweet level by also taking the time into account. We plan to study the semantic variation in language \cite{del2017semantic} in a more fine-grained way in future work.

\begin{table}[!t]
\renewcommand{\arraystretch}{1.2}
\centering
\caption{Examples of highly re-tweeted tweets after the announcement of the referendum.}
\label{tab:sarcastic}
%\footnotesize
\begin{tabularx}{\columnwidth}{X|c}\hline
\rowcolor{DGray} {\bf Tweet} & {\bf \#RT}\\\hline
\textit{They say that there is a long queue of people in ATMs but they show only 6 people waiting; this is not a queue, this is \underline{PASOK}.} &686 \\

% \textit{See this photo} (attached) \textit{so that you know who is in charge of the far-right \underline{New Democracy} of \underline{Samaras}...} (520) \\

\rowcolor{LGray} \textit{Looking for any angry tweets by \underline{SYRIZA} fans concerning Kasidiaris's} (Golden Dawn MP)\textit{ release from prison. Have you seen any?} &246\\

\textit{I want to write something funny regarding the statements made by \underline{Kammenos}} (Ind. Greeks leader)\textit{, but I cannot find something funnier than the}
\textit{statements made by \underline{Kammenos}.} &178  \\

% \textit{Now you can see why the European leaders wanted \underline{The River} to be in the government coalition.} (120) \\ 
\hline
\end{tabularx}
\end{table}

\subsection{Network}

\begin{figure}[!t]
  \centering
  \subfloat{\includegraphics[width=.85\columnwidth]{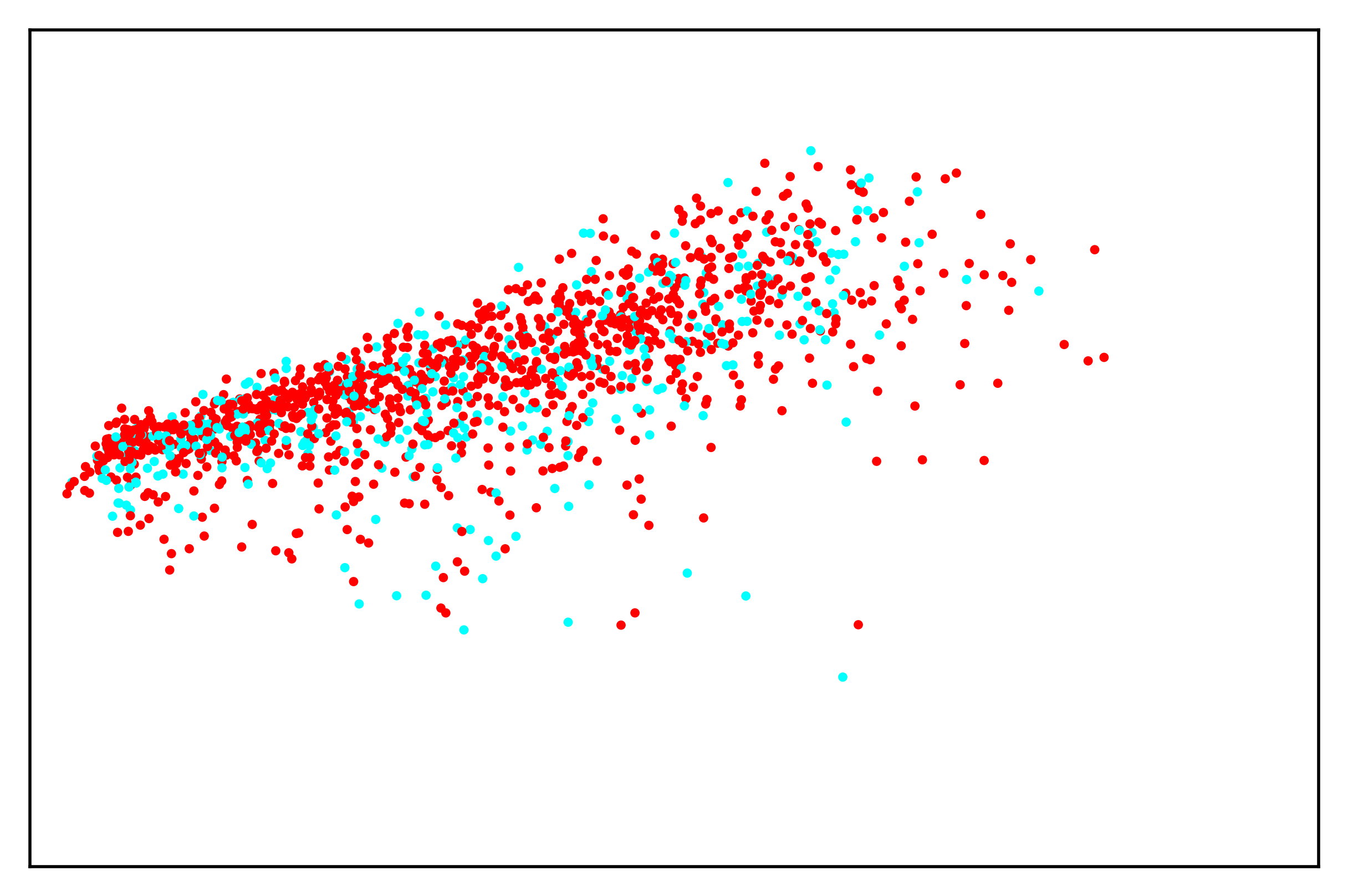}}\quad %scale=.32
  \subfloat{\includegraphics[width=.85\columnwidth]{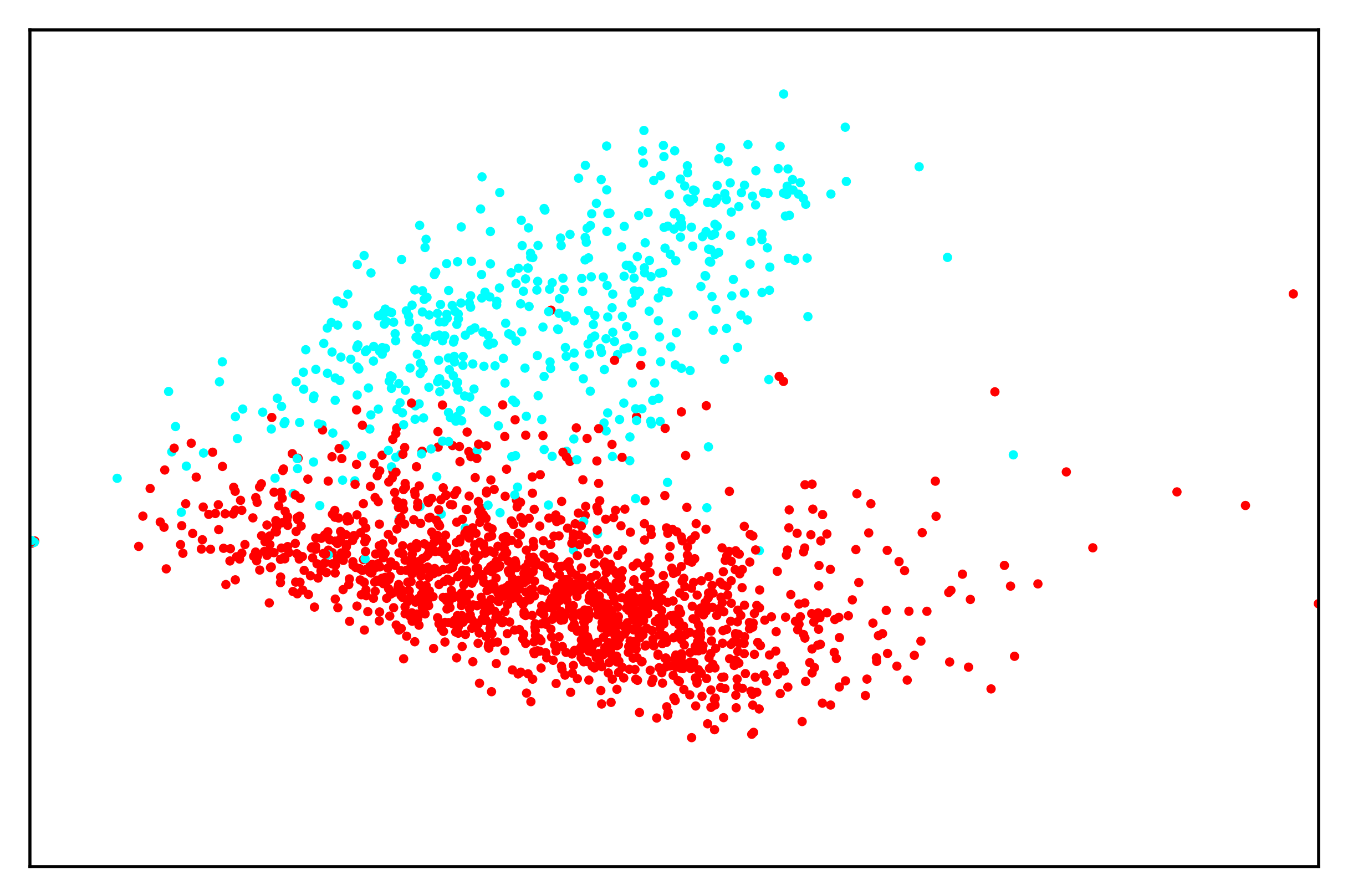}}
\caption{Network representations of \texttt{YES}/\texttt{NO} (blue/red) users, before (above) and after (below) the referendum announcement.}
\label{fig:line}
\end{figure}

We explore the differences in retweeting behaviour of users over the same periods (\textit(a) before the announcement of the referendum and  \textit(b) after and until the day of the referendum), by training two different LINE embedding models using tweets from the each period respectively. Figure~\ref{fig:line} shows the plots of the first two dimensions of the graph embeddings before and after the announcement using principal component analysis. The results unveil the effects of the referendum announcement and provide insights on the effectiveness of NETWORK information for predicting vote intention, as demonstrated in our results. Before, \texttt{YES} and \texttt{NO} users appear to have similar retweeting behaviour, which changes after the announcement. This finding illustrates the political homophily of the social network~\cite{Colleoni2014} and highlights the extremely polarised pre-election period~\cite{Tsebelis2016}.

\begin{figure}[!t]
   \centering
   \subfloat{\includegraphics[width=.48\columnwidth]{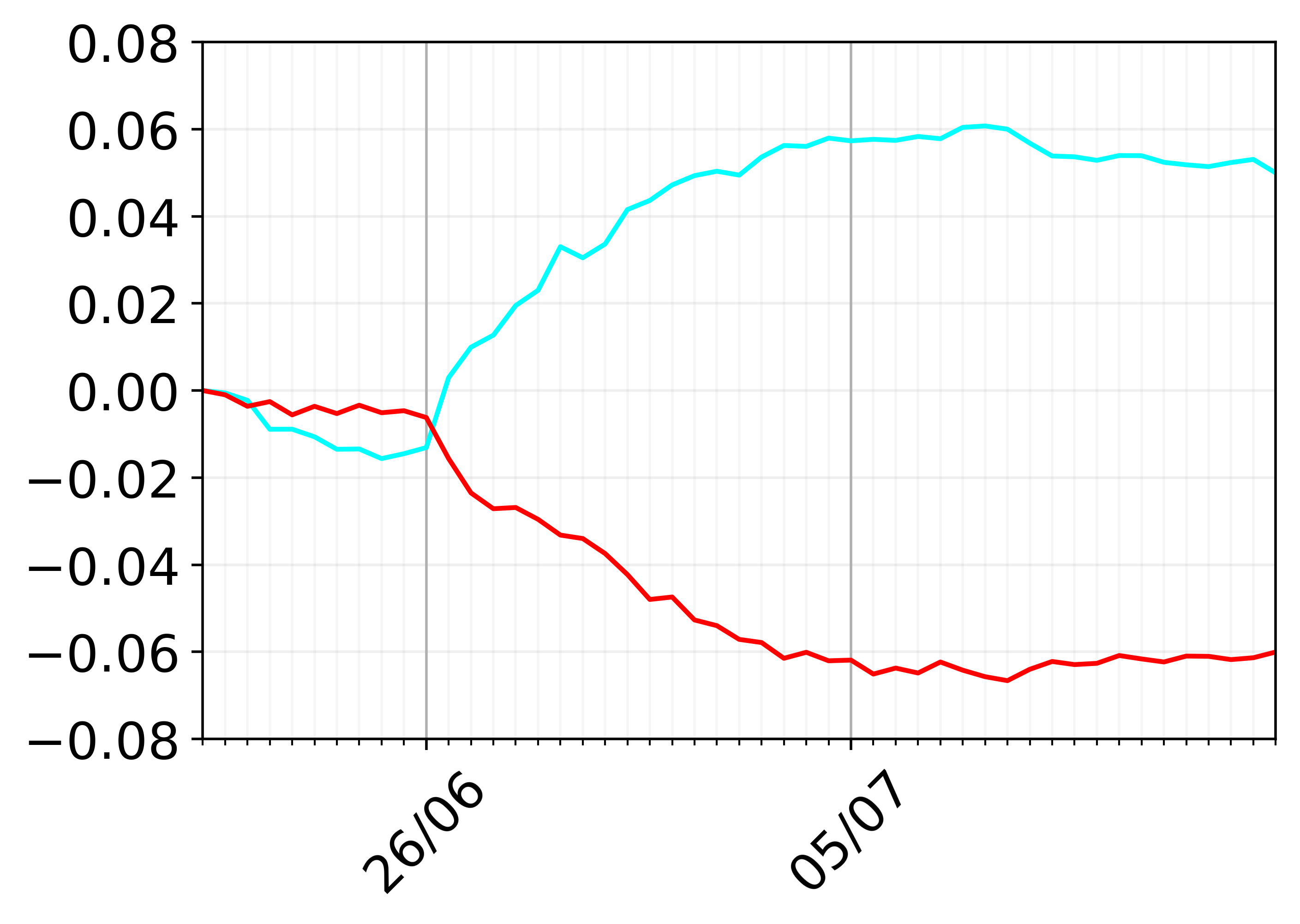}}\quad %.28
   \subfloat{\includegraphics[width=.48\columnwidth]{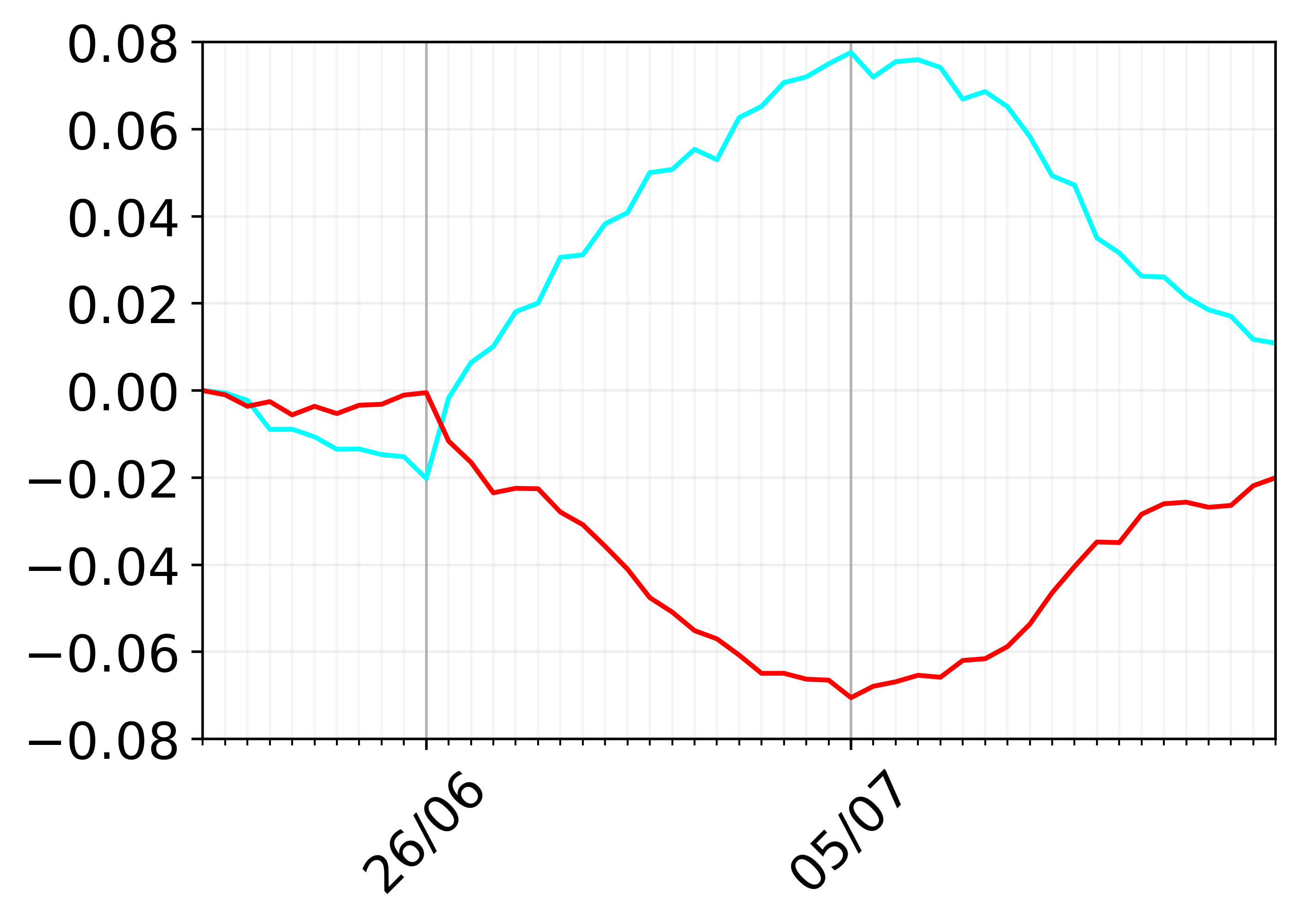}}
 \caption{Normalised difference of similarity of \texttt{YES}/\texttt{NO} (blue/red) users in our modelling (left) and in a sliding window approach (right).}
 \label{fig:line_time}
 \end{figure}

Next, we question whether the distance between the two classes of users through time changes according to time points at which real-world events occur. To answer this, we compute the network embeddings of the train and test users every 12 hours, as in our experiments, and represent every class (\texttt{YES}/\texttt{NO}) at a certain time point $t$ by the average representations ($avg_{Y}^t$, $avg_{N}^t$) of the corresponding users in the training set at $t$. Then, for every user $u$ in the test set, we use the cosine similarity $cos$ to calculate:
\begin{equation*}
    network\_score_u^t = cos(u^t, avg_{Y}^t)-cos(u^t, avg_{N}^t).
\end{equation*}
Finally, we calculate the average score of the YES and the NO users in the test set ($network_{Y}^t$, $network_{N}^t$) at every time point $t$ and normalise the corresponding time series s.t. $network_{Y}(0)$=$network_{N}(0)=0$. We also employ an alternative approach, by generating the network embeddings on a seven-day sliding-window fashion and following the same process. The results are shown in Figure ~\ref{fig:line_time}. In both cases, the \texttt{YES}/\texttt{NO} users start to deviate from each other right after the announcement of the referendum, with an upward/downward \texttt{YES}/\texttt{NO} trend until the day of the referendum. This is effectively captured in our modelling and might explain the reason for the high accuracy achieved even by our baseline models, which are trained using the network representation of the users in the last day only. However, the \texttt{YES}/\texttt{NO} users start to again approach each other only in the sliding window approach after the referendum day, since in our modelling the representations are built based on re-tweets aggregates over the whole period. While this does not seem to have affected our performance, exploring the temporal structure of the network formations through time is of vital importance for longer lasting electoral cases.

\section{Limitations and Future Work}
Despite working under a real-time simulation setting, we are aware that our results come with some caution, owed to the selection of the users in our test set. The limitations stem from the fact that we have selected highly active users that have used at least three polarised hashtags in their tweets after the announcement of the referendum. As previous work has shown \cite{cohen2013classifying,preoctiuc2017beyond}, we expect that the performance of any model is likely to drop, if tested in a random sample of Twitter users. We plan to investigate this, by annotating a random sample of Twitter users and comparing the performance in the two test sets, in our future work.

We also plan to assess the ability of MCKL to generalise, through exploring different referendum cases and incorporating more sources of information in our modelling. Finally, we plan to study the temporal variation of language and network in a more fine-grained way.

\section{Conclusion}

We presented a distant-supervised multiple convolution kernel approach, leveraging temporally sensitive language and network information to nowcast the voting stance of Twitter users during the 2015 Greek bailout referendum. Following a real-time evaluation setting, we demonstrated the effectiveness and robustness of our approach against competitive baselines, showcasing the importance of temporal modelling for our task.

In particular, we showed that temporal modelling of the content generated by social media users provides a significant boost in performance (11\%-19\% in F-score) compared to traditional feature aggregate approaches. Also, in line with past work on inferring the political ideology of social media users \cite{conover2011predicting,al2012homophily}, we showed that the network structure (in our case, the re-tweet network) of the social media users is more predictive of their voting intention, compared to the content they share. By combining those two temporally sensitive aspects (text, network) of our task via a multiple kernel learning approach, we further boost the performance, leading to an overall significant 20\% increase in F-score against the best performing, solely text-based feature aggregate baseline. Finally, we provided qualitative insights on aspects related to the shift in online discussions and polarisation phenomena that occurred during this time period, which are effectively captured through our temporal modelling approach.

\iffalse
\input{texs/1introduction.tex}
\input{texs/2backround.tex}
\input{texs/4data.tex}
\input{texs/3methods.tex}
\input{texs/5experiments.tex}
\input{texs/6results.tex}
\input{texs/7discussion.tex}
\input{texs/8conclusion.tex}
\fi
\section*{Acknowledgements}
The current work was supported by the EPSRC through the University of Warwick's Centre for Doctoral Training in Urban Science and Progress (grant EP/L016400/1) and through The Alan Turing Institute (grant EP/N510129/1).


%\bibliographystyle{ACM-Reference-Format}
%\bibliography{sample-bibliography}

\end{document}